# Virufy: A Multi-Branch Deep Learning Network for Automated Detection of COVID-19

Ahmed Fakhry[1,2], Xinyi Jiang[1], Jaclyn Xiao[1,3], Gunvant Chaudhari[1,4], Asriel Han[1,5], Amil Khanzada[1,6*]

1. Virufy AI Research Group
2. University of Alexandria, Department of Electronics and Communication
3. Duke University, Department of Biomedical Engineering
4. University of California San Francisco, School of Medicine
5. Stanford University, School of Humanities and Sciences
6. University of California Berkeley, Department of Computer Science
* Contact author: amil@virufy.ai

*Abstract*—Fast and affordable solutions for COVID-19 testing are necessary to contain the spread of the global pandemic and help relieve the burden on medical facilities. Currently, limited testing locations and expensive equipment pose difficulties for individuals trying to be tested, especially in low-resource settings. Researchers have successfully presented models for detecting COVID-19 infection status using audio samples recorded in clinical settings [5, 15], suggesting that audio-based Artificial Intelligence models can be used to identify COVID-19. Such models have the potential to be deployed on smartphones for fast, widespread, and low-resource testing. However, while previous studies have trained models on cleaned audio samples collected mainly from clinical settings, audio samples collected from average smartphones may yield suboptimal quality data that is different from the clean data that models were trained on. This discrepancy may add a bias that affects COVID-19 status predictions. To tackle this issue, we propose a multi-branch deep learning network that is trained and tested on crowdsourced data where most of the data has not been manually processed and cleaned. Furthermore, the model achieves state-of-art results for the COUGHVID dataset [16]. After breaking down results for each category, we have shown an AUC of 0.99 for audio samples with COVID-19 positive labels.

## 1 Introduction

With the spread of coronavirus disease 19 (COVID-19), over 103M positive cases have been reported by the date of February 3rd, 2021 [1]. As cases continue to fill up hospital beds in record numbers, hospitals globally have struggled to avoid reaching maximum capacity [22]. Many clinical and medical efforts have been exerted to contain the crisis, such as the creation of nucleic acid testing and amalgamating clinical characteristics of infected patients as the standard reference for detecting COVID-19 [27]. Reverse transcription-polymerase chain reaction (RT-PCR) tests are the gold standard for detecting COVID-19 in clinical practice, due to their high sensitivity and specificity [23].

Despite the reliability of RT-PCR tests, some issues have arisen during mass application. The tests require costly reagents and tools that have prevented ubiquitous access globally [26]. Furthermore, administering and processing the test comes with concern for exposure, and the test results are returned in hours to days [23]. These issues prevent RT-PCR tests from being used as primary screening for COVID-19 status. Although vaccination efforts globally are underway, distribution efforts have been impeded in low and middle income countries [25]. Many experts are uncertain about when herd immunity will be reached, especially with the emergence of new viral variants. Therefore, a fast, accurate, low-cost, and accessible screening test for COVID-19 is necessary to help limit its spread.

Emerging Artificial Intelligence (AI) technologies show promise in allowing the creation of such a solution. Deep learning and machine learning algorithms could be used to analyze cough sounds of infected patients and infer predictions. Multiple research groups have been dedicated to gathering sound recordings for COVID-19 patients of all ages, in various settings, symptomatic or asymptomatic, and at different time periods prior to symptoms onset. This ability allows AI algorithms to learn audio patterns particular to the disease for patients with different sets of demographic and medical characteristics. The most commonly collected audio recordings for detecting COVID-19 are coughs. Some groups, such as Coswara [19] and Virufy [6], collect additional counting and vowel data along with cough recordings. However, this collected data varies substantially in sound quality and formats, depending on the environment and instrument used. As there is no standard way of collecting data donor information, open and disparate data from different groups make a big portion of the metadata not suitable for training algorithms.

The state-of-the-art performance of AI algorithms for detecting COVID-19 ranges from 0.68 AUC to 0.97 AUC

[2, 15], based on several factors, such as the quality of audio files used to train the learning algorithm, preprocessing methodology, and the algorithm structure. In our previous work [6], our group developed a deep learning model with an average AUC of 0.77 across several datasets, including Coswara [19], COUGHVID [16], and Virufy [6] datasets. In this paper, we build on our previous approach and present the state-of-the-art deep learning model on the COUGHVID dataset.

We propose a multi-branch deep learning network based on fusing heterogeneous input features. Our model achieves the new State-Of-Art for the CoughVid dataset. Our proposed network successfully learns patterns from mel-frequency cepstral coefficients, mel-spectrogram images, and other clinical information to detect COVID-19 from cough audio submissions. It also distinguishes COVID-19 positive donors from healthy symptomatic donors. Our algorithm scores 0.99 AUC for detecting COVID-19 with a precision of 98.4% and a recall of 92.15% across different sections of age and gender.

## 2 BACKGROUND

Since the outbreak of COVID-19, groups like Coswara [19] and COUGHVID [16] have focused on collecting high-quality biometric data. Furthermore, research has been conducted in order to train cough-based machine learning models to detect COVID-19 through cough [5, 10, 12, 13, 15]. The high AUC exhibited by such studies [5, 13, 15] have demonstrated the increased potential of COVID-19 detection from cough data. For example, the model developed by Cambridge University had an AUC of 0.82 and used mel-frequency cepstral coefficients (MFCC) and other audio features [5]. Similarly, MIT researchers used a biomarker layer with ResNet-50 [15] based models [13] that input mel-frequency cepstral coefficients (MFCC) and output biomarkers that were used as inputs to the three parallel ResNet-50 convolutional neural networks. This model was trained on 4,256 subjects and resulted in an AUC of 0.97, highlighting the utility of AI in COVID diagnosis. Nevertheless, despite markedly high AUC values, previous models rely largely on analysis on MFCC and spectrogram inputs. Subsequently, while these models are useful in the detection of COVID, the consideration of other features may result in more clinically applicable and accurate results.

Most research on COVID-19 cough models to this date has identified only two classes for binary classification - COVID-positive or negative. While many studies have shown that cough symptoms from multiple respiratory conditions have distinguishable features [3, 9, 14, 17], COVID-19 has been noted to share similar latent cough traits as other respiratory conditions [4, 7, 18]. For instance, a patient with non-COVID-19-associated pneumonia may be falsely identified by a machine learning model as COVID positive due to the large prevalence of pneumonia that develops in COVID-positive patients [21]. Thus, classification of a patient with pneumonia as purely COVID negative may lead to erroneous identification of traits during model training.

In this paper, we propose a COVID-19 diagnostic model that integrates heterogeneous feature inputs. We adopt a three class diagnostic system previously created by Agbley et. al [2] - COVID-19 positive, symptomatic COVID-19 negative, and asymptomatic COVID-19 negative. Furthermore, by adding novel model features, we show that our model can better distinguish between COVID-19 positive patients and COVID-19 negative patients than any previous approaches, regardless of symptom status.

## 3 METHODS

### 3.1 Data

The COUGHVID dataset was utilized to train our network. As a publicly-available dataset of global cough audio recordings, COUGHVID is one of the largest COVID-19 related cough datasets. COUGHVID includes a total of 20,072 cough recordings labeled with positive COVID-19, symptomatic COVID-19 negative, and asymptomatic COVID-19 negative, along with other clinical information and metadata [16]. The dataset contains samples from a wide array of ages, genders, pre-existing respiratory conditions, and geographic locations - all of which are useful data to consider when training an unbiased deep learning model. One metadata parameter that is associated with each audio recording is the degree of certainty to which a given recording constitutes a cough sound, based on a pre-trained automatic classifier. We have filtered all audio recordings that have a degree of certainty below 0.9, which left us with 5,749 audio recordings shared among the three labels: 4,446 audio recordings for asymptomatic COVID-19 negative coughs, 923 recordings for symptomatic COVID-19 negative coughs, and 380 recordings for COVID-19 positive coughs.

### 3.2 Data Augmentation

Due to the highly imbalanced dataset, we have applied data augmentation techniques to create a more balanced training dataset. By adding gaussian noise, pitch shifting, shifting the time signal and stretching the time signal, we have increased the number of COVID-19 labeled samples from 380 to 750, along with another 750 randomly selected samples for each one of the remaining two labels. This led to a balanced dataset of 2,250 audio samples. Prior to the augmentation phase, we split the data into training, validation and test datasets such that we can apply augmentation on each split separately. By this, we were able to avoid repeated occurence of the same original audio file in two different splits to assure the validity of our network performance. For all data splits, each class is represented by one third of the number of split samples, which makes the distribution of data perfectly balanced for all classes.

|  | Training | Validation | Testing |
|---|---|---|---|
| COVID-19 positive | 600/304 | 75/38 | 75/38 |
| Symptomatic COVID-19 negative | 600/600 | 75/75 | 75/75 |
| ASymptomatic COVID-19 negative | 600/600 | 75/75 | 75/75 |

*Table 3.2: distribution of samples (samples total / samples that come from the original COUGHVID dataset) in each class in training, validation and testing split. We then use data augmentation to create samples to our desired number.*

### 3.3 Audio and Clinical Features

The way in which audio features are extracted from cough audio files has a large effect on the network performance. We train our network on mel-frequency cepstral coefficients (MFCcs) and mel-frequency spectrograms, both commonly used audio features for audio classification and speech-recognition [24]. In our network illustrated below in Fig 3.4, we use two classifiers, one of them trained on mel-spectrograms and the other on MFCCs. Each cough audio file was downsampled to half the original frequency (22KHz) and split into audio chunks. The first 13 MFCCs were extracted from the preprocessed audio chunks using the librosa package in python. The mel-spectrograms were extracted using the librosa package for the same parameters used to extract MFCCs. Each mel-spectrogram color image was reshaped to the size of (224,224,3), the original input size of the ResNet-50 convolutional neural network.

To supplement features from cough audio, we add clinical information in the COUGHVID dataset, including history of respiratory conditions and symptoms such as fever. We pass this clinical information in a one dimensional vector of binary numbers as each binary number represents the presence or absence of a symptom or a condition.

These three features extracted from each cough audio chunks are stored in a hash-table with a key for each record. The data was randomly grouped into train-validation -test sets using a 80-10-10 split.

We also did slice-based analysis and divided the test data-set into groups based on age and gender. According to age, we split the test data-set into four groups. First group is for patients below 20 years old, second is for patients between 20 & 40 years old, the third group is for patients between 40 & 60 years old while the last group is for elderly above 60 years old. For gender, we have split the test data-set into two groups, one for male and the other for females. Fig 3.3.1 and fig 3.3.2 show the histogram of the number of audio records for each group.

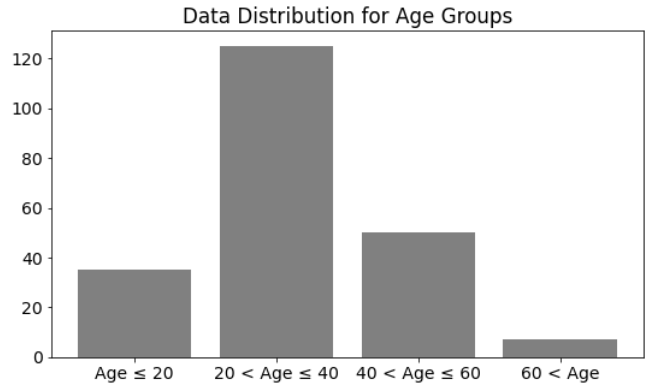

*fig 3.3.1: Histogram of number of cough audio files for each age group. The test dataset is divided into four groups based on age. The first group belongs to patients below 20 years old. The second group belongs to patients between 20 & 40 years old, while the third group belongs to patients between 40 and 60 years old. The last group belongs to elderly patients above 60 years old. The group with the most representation in the test data-set is for patients between 20 & 40 years old.*

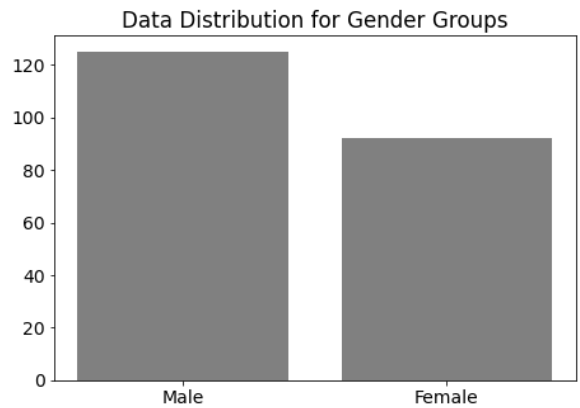

*fig 3.3.2: Histogram of number of cough audio files for each gender group. The test dataset is divided into two groups based on gender. The first group belongs to male patients while the second group belongs to female patients. Number of audio recordings are close for both groups.*

### 3.4 Model

The model is a multi-branch ensemble learning architecture based on a ResNet-50 convolutional neural network that is pre-trained on ImageNet dataset and stripped of the top layer (classification layer) [11]. The input for the CNN is a mel-spectrogram heatmap of size (224,224,3) and the output of the CNN is passed to both, a global average pooling layer and a global maximum pooling layer in two separate and parallel links. Each of them is followed by batch normalization and dropout layers before concatenated together in a single dense layer to make the first branch [fig 3.4.1].

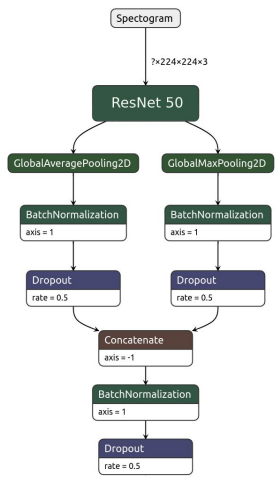

fig 3.4.1: First branch

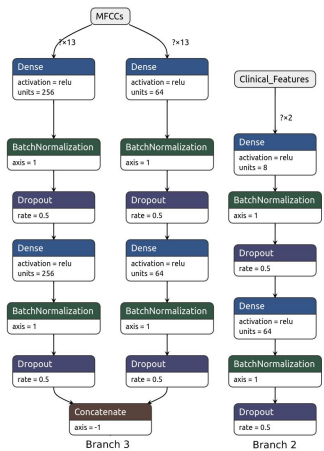

fig 3.4.2: Second branch & third branch

The second branch [fig 3.4.2] is a multi-layer feedforward neural network consisting of two dense layers that are of 8 nodes and 64 nodes, respectively. And each layer is followed by a batch normalization and dropout layers. The input for the first branch is a 1D vector of binary numbers. Each binary number represents one of the clinical information associated with the patient record, such as history of respiratory diseases, type of cough and whether the patient has fever or not. This branch is meant to incorporate the clinical information

The third branch [fig 3.4.2] is a double parallel feedforward neural network which takes a vector of mel-frequency cepstrum coefficients as an input vector of size (13,1). Each of the two parallel links is a multi-layer feed forward neural network which consists of two layers as each layer is followed by a batch normalization and dropout layers. The high ends of both links are concatenated together in a single dense layer.

The extracted high level features at the high end of the three branches are fused together before being passed to a SFFN [fig 3.4.3] that is followed by a softmax layer for a multi-label classification task. The three labels are: asymptomatic Covid-19 negative, symptomatic Covid-19 negative and positive Covid-19.

The proposed network architecture enabled us to take advantage of several heterogeneous classifiers and fuse together the extracted high-level features from spectrogram image using ResNet-50 CNN, and from MFCCs using a DNN. The network architecture, number of hidden layers for each branch and number of units per each layer were hyperparameters and were decided on using grid-search.

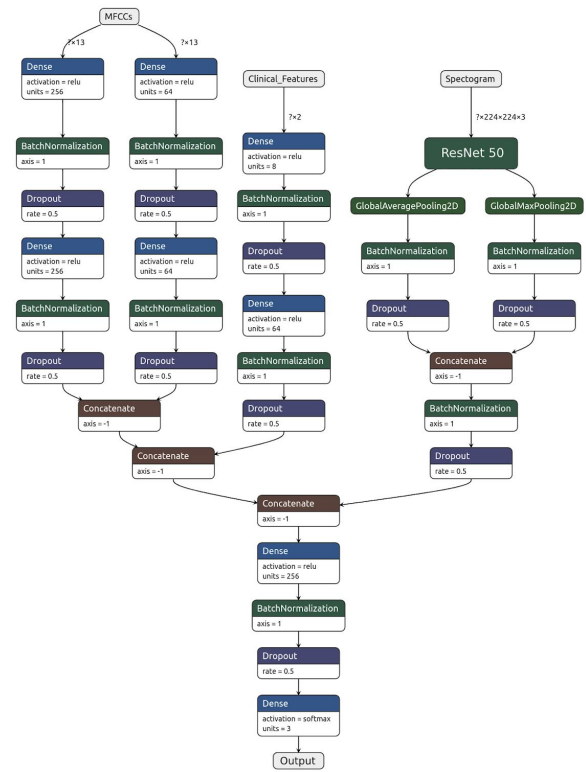

Fig 3.4.3: Multi-Branch Network Architecture. The first branch is based on a ResNet-50 convolutional neural network, the second branch is SFFN, the third branch is a DPFNN. The high ends of the three branches are concatenated together before being passed to a SFNN and a classification layer.

3.5 Statistics
Area under the ROC curve (AUC) is the main evaluation metric for the performance of our model because it represents the model performance well regardless of class balance. The AUC values for each class using one vs. all methodology and the average AUC for all classes using the micro-averaging scoring metric. We further demonstrate the validity and reliability of our model in detecting COVID-19, by calculating other statistical information, such as sensitivity, specificity, positive predictive values and negative predictive values. These statistical values are calculated for positive COVID-19 class using one vs all methodology and for a prediction threshold value of 0.9. Also, it should be noted that the occurrence of COVID-19 disease is one third in the test dataset.

## 4 RESULTS

After the success of the multi-branch model, which we have proposed in our previous work [6] with a maximum value of 0.80 AUC for COUGHVID and Coswara datasets combined, we now propose the state of the art deep learning model for COUGHVID dataset with micro-average AUC of 0.91.

Table 4.1 contains the results of our proposed network [fig 3.4] on the test dataset. Our network scores an AUC value of 0.99 for detecting positive COVID-19 coughs, implying that our model has extremely low false negative and false positive rate.

As illustrated in table 4.1, the sensitivity and specificity of detecting COVID-19 are 85% and 99.2%, respectively, which shows the high diagnostic performance of our network. In addition, despite the low prevalence of the COVID-19 disease in the test set (33.3%), the network has scored high positive-predictive value (98.4%). Such high precision and sensitivity ensures that our network will have very low false-negative results COVID-19, making it an appropriate screening for population-scale testing.

In Table 4.1, we show an ablation analysis comparing the performances of the Multi-Branch model to a single ResNet-50 model using only mel-spectrogram features. Table 4.1 contains a variety of statistical results and AUC results on separate label classes.

| Model | Class 1 AUC | Class 2 AUC | Class 3 AUC | Micro-avg |
|---|---|---|---|---|
| Multi-Branch | 0.84 | 0.82 | 0.99 | 91 |
| ResNet-50 | 0.73 | 0.69 | 0.97 | 84 |

| Model | Sensitivity | Specificity | Positive-Predictive | Negative-Predictive |
|---|---|---|---|---|
| Multi-Branch | 85% | 99.2% | 98.4% | 92.15% |
| ResNet-50 | 64% | 97.1% | 92.3% | 83.6% |

*Table 4.1: Experiment results including AUC value for each class using one vs all methodology and the average AUC for overall performance using the micro average scoring metric. Class1: asymptomatic COVID-19 negative, Class2: symptomatic COVID-19 negative, Class3: Covid-19 positive.*

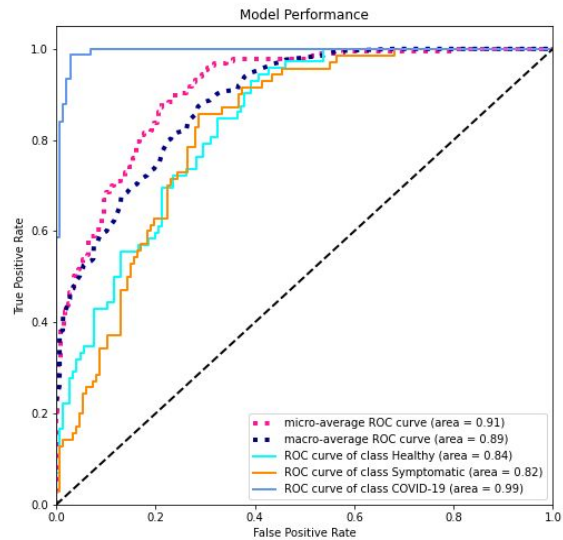

*Plot 4.2: The receiver operating characteristic curve for each class separated and calculated using one vs all methodology. Also, the micro-average and macro-average roc curves to demonstrate the diagnostic ability of the network.*

In table 4.3, we compare the performance of our network in detecting COVID-19 to another deep learning model [2] trained on the COUGHVID dataset, along with medical experts' prediction performances. Our network outperforms both experts and Agbley et al. [2] in detecting COVID-19 from cough audio files.

In table 4.4, we evaluate the performance of our network on different age groups and for both genders. We performed slice-based analysis and scored AUC values and other statistical metrics for each group. Our network successfully detects COVID-19 for all age groups with a minimum value of precision equals 91.6%, it's also equally unbiased for both genders with a minimum precision of 96%.

| Model | Sensitivity | Specificity | Positive-Predictive | Negative-Predictive |
|---|---|---|---|---|
| Multi-Branch | 85% | 99.2% | 98.4% | 92.15% |
| Experts [2] | 29% | 78% | 35% | - |
| Agbley et al. [2] | 43% | 81% | 56% | - |

*Table 4.3: A comparison in performance between our network in Agbley et al. [2], experts' prediction and the conducted model in predicting COVID-19 from cough audio samples.*

| Group (Age) | Class 1 AUC | Class 2 AUC | Class 3 AUC | Micro-avg |
|---|---|---|---|---|
| Age ≤ 20 | 0.92 | 0.90 | 1.00 | 0.95 |
| 20 < Age ≤ 40 | 0.73 | 0.75 | 0.99 | 0.83 |
| 40 < Age ≤ 60 | 0.91 | 0.83 | 1.00 | 0.94 |
| 60 < Age | 0.50 | 0.90 | 1.00 | 0.89 |
| Group (Gender) | Class 1 AUC | Class 2 AUC | Class 3 AUC | Micro-avg |
| Male | 0.77 | 0.76 | 0.99 | 0.86 |
| Female | 0.85 | 0.85 | 0.99 | 0.91 |
| Group (Age) | Sensitivity | Specificity | Positive-Predictive | Negative-Predictive |
| Age ≤ 20 | 85% | 100% | 100% | 83.33% |
| 20 < Age ≤ 40 | 75.8% | 97.9% | 91.6% | 93.06% |
| 40 < Age ≤ 60 | 72.7% | 100% | 100% | 82.3% |
| 60 < Age | 75% | 100% | 100% | 75% |
| Group (Gender) | Sensitivity | Specificity | Positive-Predictive | Negative-Predictive |
| Male | 74.4% | 98.7% | 96.9% | 88.04% |
| Female | 81.12% | 98.3% | 96.2% | 90.76% |

*Table 4.4: Slice-based analysis for different age groups and for both genders. The table includes AUC value for each class using one vs all methodology and the average AUC for overall performance using the micro average scoring metric. Class1: asymptomatic COVID-19 negative, Class2: symptomatic COVID-19 negative, Class3: Covid-19 positive. In addition to sensitivity, specificity, positive-predictive values, negative-predictive values for each group.*

## 5 CONCLUSION AND DISCUSSION

We proposed a new model architecture based on fusing heterogeneous clinical and audio features. We demonstrated an efficient way to use both MFCCs and Mel-spectrograms to train a deep neural network to detect COVID-19 status from coughs submissions. We subsequently showed that our proposed network has high performance on crowdsourced data, and compared our network performance to experts and a previously created model demonstrating high predictive values in detecting COVID-19. To ensure that our algorithm does not lose performance in demographic groups, we evaluated our detection algorithm across different slices - age groups, different geological places, and genders.

Via experiment, we have shown that the utilization of multi-branch improves the performance of the model. The three input features allow the model to capture more information and learn a variety of patterns from audio recordings and associated metadata. Each branch in the network was designed to analyze a specific type of features: ResNet-50 to extract sparse information from mel-spectrogram images, DNN to extract non-linear patterns from MFCCs, and SFFN to scale up important clinical features and symptoms such as fever and past respiratory conditions. We have shown that late fusion of the three different input features at the back-end of the three parallel branches successfully detects COVID-19 from cough audio files. We also show, in table 4.1, that our proposed network successfully distinguishes between symptomatic patients and COVID-19 patients, which opens the door for using our algorithm to detect a variety of respiratory diseases.

All training data for our model was derived from crowdsourced sources. Thus, that data's quality and character closely resemble the input to the AI system when it will be used in smartphone applications. This characteristic distinguishes our approach from previous approaches that have utilized manually cleaned data. While clean data has the potential advantage of yielding more accurate results, AI models intended for public use should be trained on the same type of potentially noisy data that is collected in practice to ensure consistent predictions.

We would also like to acknowledge a few limitations of this study. First, there has been a common issue for crowdsourcing data that it is difficult to verify the authenticity of the labels of the data. As a result, not all the labels from the COUGHVID dataset have been verified. Second, due to the difference between data collection methods and intention to maintain anonymity, COUGHVID dataset does not have some of the metadata that may be documented in other datasets [6], for example, race information. This situation might propose some difficulty as we expand our model to more datasets.

Our developed model has achieved high performance on the COVID-19 positive class coughs, proving its power to capture the viral feature in audio files. Subsequently, it can theoretically be extended to other applications such as identifying other respiratory diseases with enough patient cough samples. Furthermore, as Virufy strives to collect more data from the under-represented areas such as South America along with the ongoing process of data collection from COUGHVID [16], we plan to generalize our model to more crowdsourced datasets and improve the model performance for multiple datasets from various races, regions, and populations.

# 6 Acknowledgments

We are very grateful to Kara Meister, M.D., Stanford University Clinical Assistant Professor of Otolaryngology, and Mary L. Dunne, M.D., Stanford University Distinguished Career Institute Fellow, for their kind guidance with respect to the medical implications of our research, and Rok Sosic, Stanford University Senior AI Researcher, for his guidance on team structure and academic collaboration.